# Spatially sequential turn-on of spontaneous emission from an atomic wave packet

Marek Czachor

*Katedra Fizyki Teoretycznej i Metod Matematycznych, Politechnika Gdańska,*
*08-952 Gdańsk, ul. Narutowicza 11/12, Poland*
*e-mail: mczachor@sunrise.pg.gda.pl*

Li You

*School of Physics, Georgia Institute of Technology,*
*Atlanta, GA 30332-0430, USA*

(October 10, 2018)

We discuss a simple example demonstrating that spontaneous emission from "space-time-superposed" atomic center-of-mass wave packets is nontrivially and time-dependent modified with respect to the standard dipole-pattern typical of "space-superposed" wave packets. Our approach provides an approximate description of a nonsimultaneous interaction of electromagnetic field with different parts of a wave packet.

## I. INTRODUCTION

Standard treatments of spontaneous emission from atomic wave packets (Cohen-Tannoudji *et al.*, 1991; Graham *et al.*, 1992; Rząźewski and Żakowicz, 1992; Ren *et al.*, 1992 Rząźewski *et al.*, 1994; Steuernagel and Paul, 1995; Stoop and Rząźewski, 1995) assume situations where the atomic exited state couples to the electromagnetic vacuum at all points of the wave packet "at $t = 0$". For $t > 0$ the atom starts to decay. The beginning of this process is assumed to take place at all points of the wave packet simultaneously. Formally this means that at $t = 0$ the wave packet is in a superposition of the kets $|r, +\rangle$, where $r$ is an eigenvalue of the atomic center-of-mass position operator, and "+" means an electronically excited state. What happens for $t < 0$ is usually ignored.

An experimental procedure that leads to such a situation may be based on a simultaneous excitation of the atom at all points occupied by the wave packet. Alternatively, one can think of situations where a metastable long-lived state is coupled to the vacuum reservoir by a simultaneous turn-on of an electric field in the whole region where the atom is supposed to reside. It is clear that the outlined scenario (below referred to as a "simultaneous spontaneous emission" (SSE)) does not cover all the experimental possibilities. First of all, the atomic wave packets can be *macroscopically* extended. This happens practically in all atom interferometer experiments where a phase shift is produced by applying different fields to different parts of the interfering wave packets. Although typical experiments involve *static* arrangements there is basically no problem with manipulating the atoms in a fully controllable time-depended manner. One can imagine, for example, a situation where different parts of the wave packet are either *nonsimultaneously excited* or *nonsimultaneously couple* to the vacuum. Actually, at the moment of writing the paper we are aware of an experiment where the latter possibility was practically realized. What is interesting the experimental data do not completely agree with the standard SSE computations.

The works of Cohen-Tannoudji *et al.* (1991) and Graham *et al.* (1992) suggest that either there is completely no influence of the coherence of an atomic wave packet on the radiated photons, or that there is at least no coherence in the emitted light if the atomic wave packets do not overlap. The work of Rząźewski and Żakowicz (1992), on the other hand, shows clearly that spectral density of radiation carries some information about the shape of the wave packet even if one considers a two-peaked Gaussian whose peaks are separated by several wavelengths of photons and are essentially non-overlapping. The two-peaked wave function in position space translates into an oscillatory wave function in momentum space, therefore manifesting the modified probability distribution of momenta in the wave packet. This momentum distribution implies a distribution of Doppler shifts in the spectral density of emitted light. This interesting result means that, in principle, one can investigate a structure of atomic wave packets by detecting photons emitted during such an apparently incoherent process as spontaneous emission.

In the discussed experiment (Robert *et al.*, 1992) a two-peaked hydrogen wave packet was prepared in an atomic interferometer in the internal metastable $2s_{\frac{1}{2}}$ state, and arranged to enter a region of space with an electric field where the Stark mixing between the $2s_{\frac{1}{2}}$ and $2p_{\frac{1}{2}}$ states led subsequently to spontaneous emission of a Lyman-$\alpha$ photon. Since the atom was moving, different parts of the wave packet were coupled to the field at different times and thus the spontaneous decay was sequentially turned on. The nonsimultaneity of the coupling was not taken into account by Rząžewski and Żakowicz (1992) which may be a source of the disagreement between the theory and the experiment.

A nonsimultaneous coupling between the field and the atom will generally lead to a complicated, infinite set of coupled integro-differential equations that are quite difficult to solve even numerically. We have decided therefore to consider a problem which on one hand is simpler and on the other exhibits physical properties which are expected to occur if a nonsimultaneous coupling is involved. The results we discuss cannot be directly compared to the experimental data for two reasons. First of all, to simplify numerical computations we shall assume that the wave packet is a one-peaked and not a two-peaked Gaussian. Second, we model the nonsimultaneity by considering an ideal situation where the decay is turned on by a moving step function. The experiment was not that ideal and the sequential turn-on was much less abrupt.

Nevertheless, in spite of these simplifications, the result we report is highly nontrivial. It shows that the "nonsimultaneous spontaneous emmission" (NSSE) is drastically modified with respect to the SSE case. The radiation pattern exhibits a time-dependent anisotropy which depends on details of the turn-on. In the infinite velocity or infinite time limits it reduces to the ordinary Doppler-modified dipole pattern of Rząžewski *et al.*.

## II. SPONTANEOUS EMISSION FROM A "SPACE-TIME SUPERPOSED" ATOMIC WAVE PACKET

The standard analysis of SSE involves "equal-time and spacelike" superpositions of the atomic center-of-mass eigenstates. In what follows we shall consider superpositions of the center-of-mass eigenstates which additionally start to decay in a position-dependent way. One can say that the choice of the "$t = 0$" varies from point to point. We will concentrate on the simple case involving a single Gaussian whose decay is turned on by a propagating step function moving in the $-z$ direction with a constant velocity $v$. Some preliminary results of such computations were briefly discussed in Czachor and Li (1996). More complicated examples, including the case of a "beaded" two Gaussian wave packet, will be discussed elsewhere.

We begin with the nonrelativistic dipole (RWA) Hamiltonian

$$H = \frac{\bm{p}^2}{2M} + H_A + H_F + H_{AF}, \tag{1}$$

with

$$\begin{aligned} H_A &= \frac{1}{2}\hbar\omega_0\sigma_3, \\ H_F &= \sum_{\bm{k},\mu} \hbar\omega_{\bm{k}} a^\dagger_{\bm{k}\mu} a_{\bm{k}\mu}, \\ H_{AF} &= -i\sum_{\bm{k},\mu} \hbar g_{\bm{k}\mu} e^{i\bm{k}\cdot\bm{r}} \sigma_+ a_{\bm{k}\mu} + \text{h.c.}, \end{aligned} \tag{2}$$

which are kinetic, internal, free field, and dipole interaction parts respectively. The dipole coupling strength is $g_{\bm{k}\mu} = \mathcal{E}_{\bm{k}}\wp(\bm{e}_\wp \cdot \bm{e}_{\bm{k}\mu})/\hbar$, where $\bm{e}_\wp$ is a unit vector of the dipole operator, and $\wp$ is the dipole moment of the electronic transition; $\bm{e}_{\bm{k}\mu}$, $\mu = 1, 2$, are the photon polarization vectors, and $\mathcal{E}_{\bm{k}} = \sqrt{2\pi\hbar\omega_{\bm{k}}/\mathcal{V}}$ is the electric field strength per photon in the quantization volume $\mathcal{V}$.

We assume the state of the atom-field system to be

$$\begin{aligned} |\psi_S\rangle &= \int d\bm{r}\, \alpha_t(\bm{r})|\bm{r}; +; 0\rangle \\ &+ \sum_{\bm{k},\mu} \int d\bm{r}\, \beta_t(\bm{r}, \bm{k}, \mu)|\bm{r}; -; \bm{k}, \mu\rangle, \end{aligned} \tag{3}$$

where

$$\begin{aligned} \alpha_t(\bm{r}) &= \big(\frac{1}{2\pi\hbar}\big)^{\frac{3}{2}} \int d\bm{p}\, \alpha_t(\bm{p}) e^{i\bm{p}\cdot\bm{r}/\hbar}, \\ \beta_t(\bm{r}; \bm{k}, \mu) &= \big(\frac{1}{2\pi\hbar}\big)^{\frac{3}{2}} \int d\bm{p}\, \beta_t(\bm{p}; \bm{k}, \mu) e^{i(\bm{p}-\hbar\bm{k})\cdot\bm{r}/\hbar}, \end{aligned} \tag{4}$$

with $\bm{p}$ and $\bm{r}$ being the center-of-mass momentum and position, "+" ("−") correspond to excited (ground) internal atomic state, and $\bm{k}$, $\mu$ denote momentum and polarization of the emitted photon.

The solution of the above model for the case of SSE is given by (Rząźewski and Żakowicz, 1992)

$$\alpha_t(\boldsymbol{p}) = \alpha_0(\boldsymbol{p})e^{-s_0 t\Theta(-t)-(\gamma+s_0)t\Theta(t)},$$

$$\beta_t(\boldsymbol{p},\boldsymbol{k},\mu) = -\alpha_0(\boldsymbol{p})g_{\boldsymbol{k}\mu}\frac{e^{-(\gamma+s_0)t\Theta(t)} - e^{-s_k t\Theta(t)}}{\gamma + s_0 - s_k}$$

$$= -\Theta(t)\alpha_0(\boldsymbol{p})g_{\boldsymbol{k}\mu}\frac{e^{-(\gamma+s_0)t} - e^{-s_k t}}{\gamma + s_0 - s_k} \tag{5}$$

with

$$s_0 = \frac{i}{\hbar}\left(\frac{\boldsymbol{p}^2}{2M} + \frac{1}{2}\hbar\omega_0\right),$$

$$s_k = \frac{i}{\hbar}\left[\frac{(\boldsymbol{p}-\hbar\boldsymbol{k})^2}{2M} - \frac{1}{2}\hbar\omega_0 + \hbar\omega_{\boldsymbol{k}}\right], \tag{6}$$

where $\gamma$ determines the lifetime. The step function $\Theta(t)$ is 1 for $t \geq 0$ and 0 otherwise. The $\Theta(-t)$ in the exponents in (5) represents the fact that the evolution of the wave packet for $t < 0$ is free and uncoupled to the vacuum field reservoir. To generate the NSSE solution we superpose the center-of-mass eigenstates with amplitudes (4) but now instead of $\Theta(t)$, which is responsible for the simultaneous turn-on, we take a more general superposition with amplitudes containing $\Theta(f(t,\boldsymbol{r}))$. The "switch-on function" $f(t,\boldsymbol{r})$ depends on details of the experimental arrangement. Consider

$$|\psi_{NS}\rangle = \int d\boldsymbol{r}\, \alpha_{t,v}(\boldsymbol{r})|\boldsymbol{r};+;0\rangle$$

$$+ \sum_{\boldsymbol{k},\mu}\int d\boldsymbol{r}\, \beta_{t,v}(\boldsymbol{r},\boldsymbol{k},\mu)|\boldsymbol{r};-;\boldsymbol{k},\mu\rangle, \tag{7}$$

where $\alpha_{t,v}(\boldsymbol{r}) = \alpha_{t+z/v}(\boldsymbol{r})/\|\psi_{NS}\|$ and

$$\beta_{t,v}(\boldsymbol{r},\boldsymbol{k},\mu) = \beta_{t+z/v}(\boldsymbol{r};\boldsymbol{k},\mu)/\|\psi_{NS}\| \tag{8}$$

$$= -\frac{1}{\|\psi_{NS}\|}\left(\frac{1}{2\pi\hbar}\right)^{\frac{3}{2}}\int d\boldsymbol{p}\, e^{i(\boldsymbol{p}-\hbar\boldsymbol{k})\cdot\boldsymbol{r}/\hbar}\alpha_0(\boldsymbol{p})g_{\boldsymbol{k}\mu}\frac{e^{-(\gamma+s_0)(t+z/v)\Theta(t+z/v)} - e^{-s_k(t+z/v)\Theta(t+z/v)}}{\gamma + s_0 - s_k} \tag{9}$$

$$= -\frac{1}{\|\psi_{NS}\|}\left(\frac{1}{2\pi\hbar}\right)^{\frac{3}{2}}\Theta(t+z/v)e^{-i\boldsymbol{k}\cdot\boldsymbol{r}}\int d\boldsymbol{p}\, e^{i\boldsymbol{p}\cdot\boldsymbol{r}/\hbar}\alpha_0(\boldsymbol{p})g_{\boldsymbol{k}\mu}\frac{e^{-(\gamma+s_0)(t+z/v)} - e^{-s_k(t+z/v)}}{\gamma + s_0 - s_k}. \tag{10}$$

The state (7) corresponds to the process where the emission is turned on by the step function $\Theta(t+z/v)$ propagating in the $-z$ direction with velocity $v$. (This should not be confused with a description in a moving frame!) Obiously

$$\beta_{t,\infty}(\boldsymbol{r},\boldsymbol{k},\mu) = \beta_t(\boldsymbol{r},\boldsymbol{k},\mu) \tag{11}$$

as it should be. The $t \to \infty$ limit of (7) will not differ from the analogous simultaneous case since the contributions from $\Theta(t(1+z/(tv)))$ for $t \to \infty$ and $v$ fixed are important only for the $z \to \infty$ parts of the wave packet which are assumed to vanish. Nevertheless, this asymptotic behavior does not mean that the problem completely reduces to the simultaneous case. The transient modifications of the angular distribution shown in Fig. 2 are relatively long-lived and in principle can last arbitrarily long, depending on relative scales of $\tau_{\text{natural}}$, $v$ and the size of the wavepacket.

Denote $\tau = \tau(t,z) = t + z/v$. The probability density for NSSE of emitting a photon with momentum $\boldsymbol{k}$ and polarization $\mu$ is

$$\mathcal{P}_{t\boldsymbol{k}\mu} = \int d\boldsymbol{r}|\beta_{t,v}(\boldsymbol{r},\boldsymbol{k},\mu)|^2 = \frac{|g_{\boldsymbol{k}\mu}|^2}{2\pi\hbar\|\psi_{NS}\|^2}\int d\boldsymbol{p}\int d\boldsymbol{p}'\int dz\,\Theta(\tau)$$

$$\exp\left[-2\gamma\tau + \frac{i}{\hbar}\left[(p_z - p'_z)z + \frac{p_z^2 - p'^2_z}{2M}\tau\right]\right]$$

$$\alpha_0(\boldsymbol{p})\alpha_0^*(p_x, p_y, p'_z)\mathcal{L}(\boldsymbol{p})\mathcal{L}^*(p_x, p_y, p'_z), \tag{12}$$

with

$$\mathcal{L}(\boldsymbol{p}) = \frac{1 - \exp\left[i(\frac{p_x k_x}{M} + \frac{p_y k_y}{M} + \frac{p_z k_z}{M} - \delta_k - i\gamma)\tau\right]}{\frac{p_x k_x}{M} + \frac{p_y k_y}{M} + \frac{p_z k_z}{M} - \delta_k - i\gamma}, \tag{13}$$

where we have defined $\delta_k = \omega_k + \hbar\omega_k^2/(2Mc^2) - \omega_0$. Without loss of generality, we can put $k_y = 0$, (which simply specifies the coordinates). In this work, we assume the wave packet is initially coherent and has a Gaussian distribution in momentum space, given by

$$\alpha_0(\boldsymbol{p}) = \left(\frac{a}{\hbar\sqrt{2\pi}}\right)^{\frac{3}{2}} \exp\left(-\frac{p^2 a^2}{4\hbar^2}\right). \tag{14}$$

In this case the integration over $p_y$, $p_z$ and $p_z'$ can be performed to obtain

$$\mathcal{P}_{t\boldsymbol{k}\mu} \propto \int dp_x \int dz \Theta(\tau) \exp\left(-2\gamma\tau - \frac{p_x^2 a^2}{2\hbar^2}\right) |F|^2, \tag{15}$$

with

$$F(z, p_x) = \frac{i\pi M}{|k_z|} \left[\exp\left(\frac{b_1^2}{a_t^2}\right) \mathcal{W}(\eta_1) - e^{-i\Delta_k \tau} \exp\left(\frac{b_2^2}{a_t^2}\right) \mathcal{W}(\eta_2)\right], \tag{16}$$

where $\mathcal{W}(\xi) = \exp(-\xi^2)\mathrm{erfc}(-i\xi)$ is the exponentially scaled error function, and

$$\begin{aligned}
b_1 &= iz, \\
b_2 &= i\left(z + \frac{\hbar k_z}{M}\tau\right), \\
\Delta_k &= \delta_k + i\gamma - \frac{p_x k_x}{M}, \\
\eta_j &= \frac{Ma_t}{2\hbar|k_z|}\left(\Delta_k - b_j \frac{2\hbar k_z}{Ma_t^2}\right), \qquad j = 1, 2.
\end{aligned} \tag{17}$$

The time-dependent width of the wave packet (due to spreading) is $a_t^2 = a^2 - 2i\hbar\tau/M$. In the asymptotic limit of large $t$, only the second term in Eq. (16) survives. Using the asymptotic expansion $\sqrt{\pi}\eta\mathcal{W}(\eta) \sim i$ for $\eta \to \infty$, we can show that our general result for (19) reduces to the SSE case (Rzążewski and Żakowicz, 1992) if $v \to \infty$.

The total probability of spontaneous emission is now

$$\sum_{\boldsymbol{k}\mu} \mathcal{P}_{t\boldsymbol{k}\mu} = \int d\Omega_{\boldsymbol{k}} (1 - |\boldsymbol{e}_{\wp} \cdot \boldsymbol{e}_{\boldsymbol{k}}|^2) P_t(\theta), \tag{18}$$

with $P_t(\theta)$ the reduced probability distribution that is defined as

$$\begin{aligned}
P_t(\theta) &= \int d\omega_{\boldsymbol{k}} \mathcal{Q}_t(\omega_{\boldsymbol{k}}), \\
\mathcal{Q}_t(\omega_{\boldsymbol{k}}) &= \frac{3\gamma a^2}{16\pi^3 \hbar^3 \parallel \psi_{NS} \parallel^2} \int dp_x \int dz \Theta(\tau) \\
&\quad \exp\left(-2\gamma\tau - \frac{p_x^2 a^2}{2\hbar^2}\right) |F(z, p_x)|^2,
\end{aligned} \tag{19}$$

and $\theta$ is the polar angle between $\boldsymbol{k}$ and the z-axis.

For the case of hydrogen $2p_{\frac{1}{2}} \to 1s_{\frac{1}{2}}$ transition, the parameters are $\gamma = (2\pi)50$ MHz, $\lambda_0 = 121.6$ nm, the recoil velocity $v_{\mathrm{recoil}} = \hbar k_0/M = 3.25$ (m/s) and energy $\omega_{\mathrm{recoil}} = \hbar k_0^2/(2M) = (2\pi)13.328$ MHz.

In Fig. 1, the results of the time-dependent (normalized) spectrum are shown for the case of $v = 10(v_{\mathrm{recoil}})$ and $v = 1(v_{\mathrm{recoil}})$. We have used $a = \lambda_0$ as normally assumed for a cold atomic wave packet in the recoil limit. With these parameters, the time $a/v$ it takes to traverse the wave packet is much less than the lifetime of the metastable $2s_{\frac{1}{2}}$ state, therefore neglecting its decay is a well justified approximation. We see that due to the time dependence introduced by the swept turning on process, a general broadening of the spectrum is seen, but in the limit $t \to \infty$, it approaches the result known for the case of the SSE. This time-dependent broadening can be easily understood if one considers a wave packet which is totally localized in a region of space where the step function has not yet arrived. The spectrum must then be completely flat (equal identically to 0) since the atom cannot radiate. However, after the exitation (or coupling) is completed the atom will radiate and the radiation pattern must be nonzero. Our results show that the pattern continuously deforms from a flat distribution to a Lorentzian. The dependence on the polar angle $\theta$ exists as well but the results will not be explicitly shown here. Instead we show in Figs. 2-3 the total integrated signal (over the whole spectrum range) of the reduced (i.e. divided by the dipole pattern) probability distribution. This quantity should be constant in the case of a simultaneous spontaneous emission and is clearly modified by the non-simultaneity of "excitation". The effect is transient and should asymptotically approach the dipole distribution for very large $t$. Note nevertheless that the range of time we consider at Fig. 2 is relatively large in atomic scales ($-50\tau_{\text{natural}} < t < +40\tau_{\text{natural}}$ in Fig. 2). For the wave packet we consider the effect should be observable for a few hundred of natural lifetimes and can be more long-lived for different wave packets and excitation velocities.

The results suggest that when the time of transit of the "excitation" through the wave packet is of the order of the lifetime of the excited state, the properties of the spontaneous emission may be significantly modified. One may speculate that this kind of modification may be the source of the "fringes" observed by Robert *et al.* (1992).

To complete the analysis let us mention that we have taken the initial wave packet $\alpha_0(\boldsymbol{r})$ to be completely coherent. A more general treatment would require treating both $\alpha_t(\boldsymbol{r})$ and $\beta_t(\boldsymbol{r})$ as statistical wave functions starting with a partially coherent $\alpha_0(\boldsymbol{r})$ described by the spatial coherence function $\Gamma(\boldsymbol{r}, \boldsymbol{r}') = \langle \alpha_0(\boldsymbol{r})\alpha_0(\boldsymbol{r}')\rangle$. Our result can be easily extended to the case of a partially coherent initial wave packet provided the coherence function $\Gamma(\boldsymbol{r}, \boldsymbol{r}')$ is given. Depending on the particular form of $\Gamma(\boldsymbol{r}, \boldsymbol{r}')$, the properties of the spontaneous emission as given by the solution (12) may be different. The richness of the structure we observe in such a simple case suggests that the approach developed here may allow for a determination of the partial coherence function $\Gamma(\boldsymbol{r}, \boldsymbol{r}')$ by studying its NSSE. In a wider perspective, we can speculate that this relation is not only restricted to the case of spontaneous emissions. It should also be true for any other kind of non-simultaneous interactions, such as scattering of light from different parts of a wave packet at different times. In view of the recent success in achieving the Bose-Einstein condensation of alkali atoms (Anderson *et al.*, 1995; Davies *et al.*, 1995; Andrews *et al.* 1997; Mewes *et al.*, 1997) we expect such a kind of non-simultaneous interaction scheme could provide a possible diagnostic scheme for detecting the partial coherence function, and revealing the off diagonal long range order of the condensate.

## III. FINAL REMARKS

A closer look at the SSE solutions given in Rząźewski and Żakowicz (1992) shows that they very well agree with an intuitive picture of what is going on: We simply obtain a coherent superposition of the atom-photon states representing Doppler shifted emitted photons. The probability distribution of the atomic momenta manifests itself in an analogous distribution of the frequency shifts with respect to the dipole pattern. This explains the oscillatory behaviour of the spectral density of radiation which is completely unrelated to any photon interference. This is consistent with what we know about entanglement of quantum states and its role for a two-particle interference (Czachor, 1991). What is encourageing, however, is the fact that one could obtain an identical result without any center-of-mass calculations but simply by taking a superposition of suitably Doppler shifted Weisskopf-Wigner wave functions.

This observation motivated the analysis presented in this Letter. We simply analogously superpose the Weisskopf-Wigner wave functions but in addition to the Doppler and recoil shifts we take into account the fact that parts of the wave packet which are "excited" at different times must have the initial time-of-decay "$t=0$" different for different points of the wave packet.

One may wonder whether on general grounds the approach is acceptable. For example, the moving step-function-excitation involves an interaction of the initial wave packet with some external fields. As such, it may lead to a decoherence of the atomic state. The decoherence-recoherence experiment reported by the MIT group (Chapman *et al.*, 1995) shows however that an atomic wave packet may remain coherent after having interacted with a laser beam and even having spontaneously emitted a photon provided one afterwards purifies the state in a suitable correlation experiment.

To close these remarks let us note that even the SSE is simultaneous in only one reference frame. And even though the nonsimultaneity experienced by a moving observer is not exactly of the kind we discuss in this Letter, it shows that superpositions of states corresponding to "different positions and different times of events" are physically unavoidable.

## ACKNOWLEDGMENTS

M. C. is grateful to Prof. Kazimierz Rzążewski who introduced him to the problem and pointed out the importance of nonsimultaneity of excitation for a correct interpretation of such experiments, and to Prof. David E. Pritchard for hospitality in his lab, help and many hours of fruitful discussions. We thank all members of the MIT and Paris-Nord atom interferometer groups for their interest in this work and friendly atmosphere they created. L. Y. wants to thank Prof. A. Starace for helpful discussions and critical reading of the manuscript. The work of L. Y. was supported by the National Science Foundation through a grant for the Institute for Theoretical Atomic and Molecular Physics at Harvard University and Smithsonian Astrophysical Observatory. The work of M. C. was supported by the Fulbright Commission and the Polish-Flemish grant 007.

[1] Anderson, M. H., Ensher, J. R., Matthews, M. R., C. E. Wieman, C. E., and Cornell, E. A. (1995). *Science* **269**, 198.
[2] Andrews, M. R. *et al.* (1997). Science **275**, 637.
[3] Bradley, C. C., Sackett, C. A., Tollett, J. J., and Hulet, R. G., (1995). *Physical Review Letters*, **75**, 1687.
[4] Chapman, M. S., Hammond, T. D., Lenef, A., Schmiedmayer, J., Rubenstein, R. A., Smith, E., and Pritchard, D. E. (1995). *Physical Review Letters*, **75**, 3783.
[5] Cohen-Tannoudji, C., Bardou, F. and Aspect, A. (1991) *Laser Spectroscopy X*, ed. by M. Ducloy *et al.*, World Scientific.
[6] Czachor, M. (1991). Foundations of Physics Letters **4**, 351.
[7] Czachor, M. and You, L. (1996). *Coherence and Quantum Optics VII*, edited by J. H. Eberly, L. Mandel, and E. Wolf, Plenum Press, New York.
[8] Davies, K. B., Mewes, M.-O., Andrews, M. R., van Druten, N. J., Durfee, D. S., Kurn, D. M., and Ketterle, W. (1995). *Physical Review Letters* **75**, 3969.
[9] Graham, R., Walls, D. F., and Zoller, P. (1992). *Physical Review* A **45**, 5018.
[10] Mewes, M.-O. *et al.* (1997). Physical Review Letters, **78**, 582.
[11] Ren, W., Cresser, J. D., and Carmichael, H. J. (1992). *Physical Review* A **46**, 7162.
[12] Robert, J., Miniatura, Ch., Gorceix, O., Le Boiteux, S., Lorent, V., Reinhardt, J., and Baudon, J. (1992). *Journal de Physique II*, **2**, 601.
[13] Rzążewski, K. and Żakowicz, W. (1992). *Journal of Physics* B **25**, L319.
[14] Rzążewski, K. *et al.* (1994). *Acta Physica Polonica* A, **86**, 279.
[15] Steuernagel, O. and Paul, H. (1995). *Physical Review* A **52**, R905.
[16] Stoop, H. and Rzążewski, K. (1995). *Physical Review* A **52**, 1494.

FIG. 1. Normalized spectrum ($\theta = 0$) at three different times corresponding to the edge of the moving theta function at $z = 0.5\lambda_0$ (dotted line), $z = 0$ (solid line), and $z = -0.5\lambda_0$ (dashed line). Dash-dotted lines show the unmodified Lorentzian natural lineshape. The step function moves along the $z$ axis with velocity $v$ in the negative direction (from right to left).

FIG. 2. The reduced angular distribution for $v = 1(v_{\text{recoil}})$ with other parameters the same as in Fig. 1. as a function of time and $\theta$. At $t = 0$ the step function arrives at the center of the Gaussian. An analogous plot is completely flat in the SSE case.

FIG. 3. The reduced angular distribution of radiation at the time corresponding to the edge of the moving theta function at $z = -0.5\lambda_0$ for three different velocities (in units of $v_{\text{recoil}}$). The rest of parameters the same as in Fig. 1. For $v = v_{\text{recoil}}$ the atom tends to radiate in the direction of propagation of the "excitation". For $v = 0.1 v_{\text{recoil}}$ the radiation is concentrated in the direction perpendicular to the excitation. With growing velocity of the excitation the distribution approaches the dipole pattern.

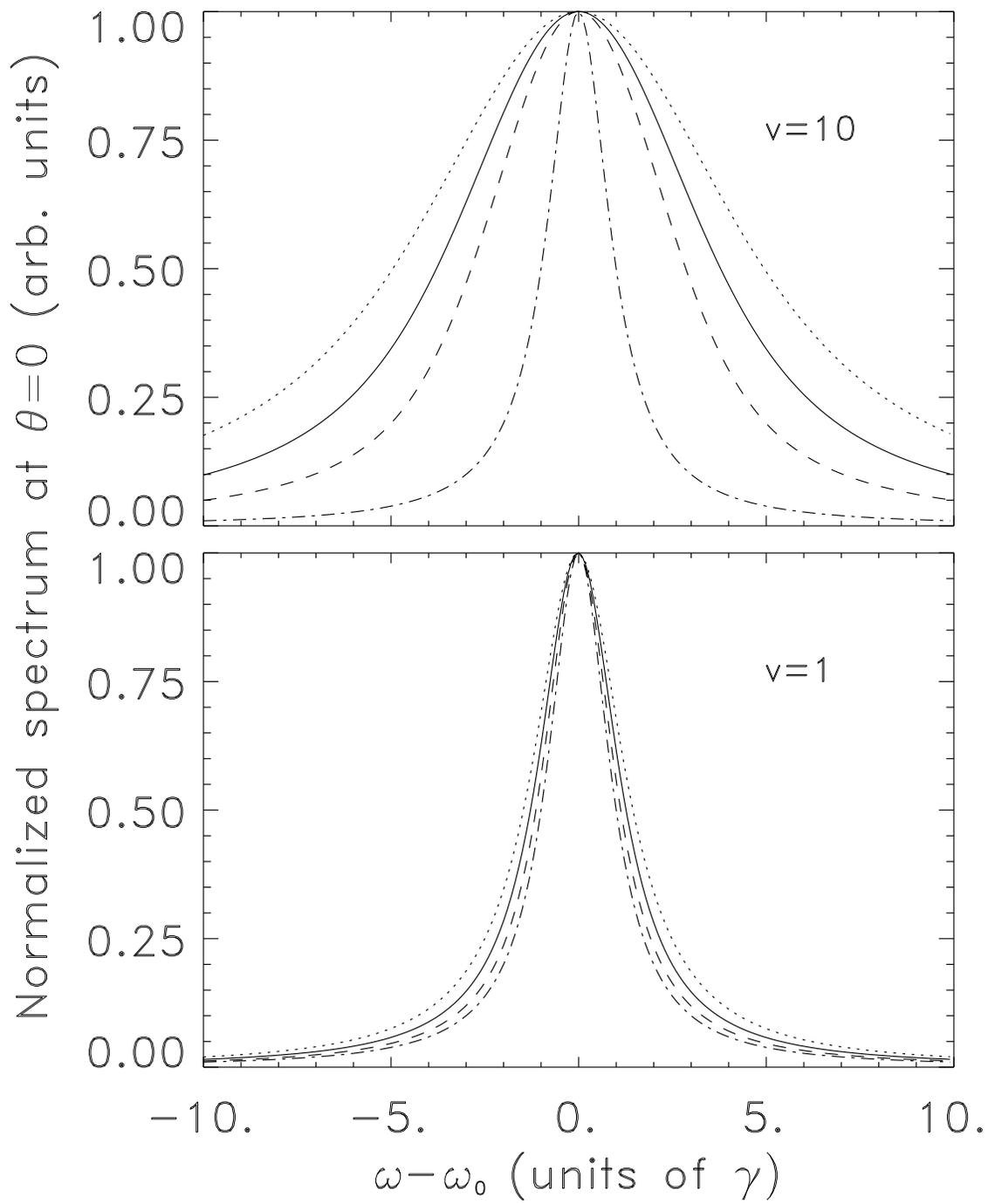

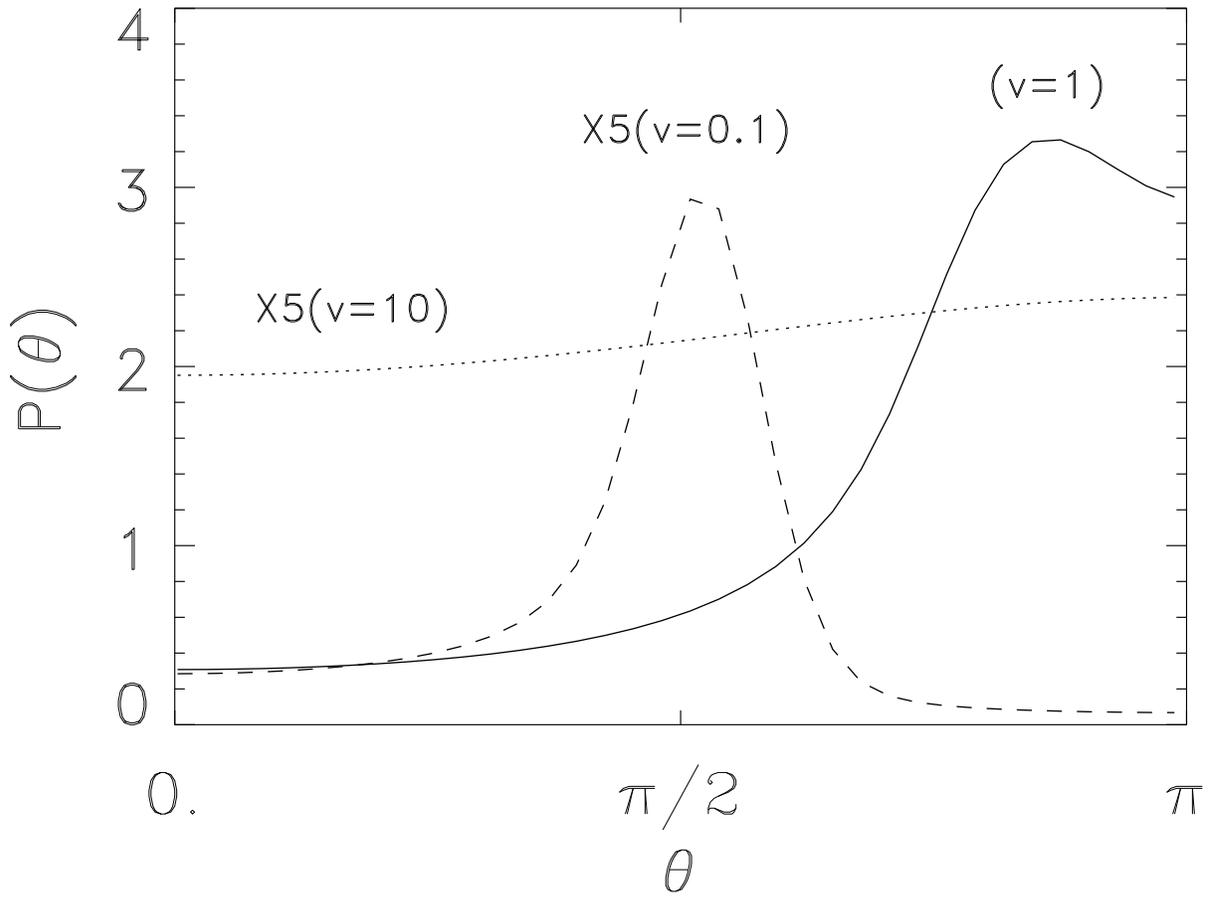

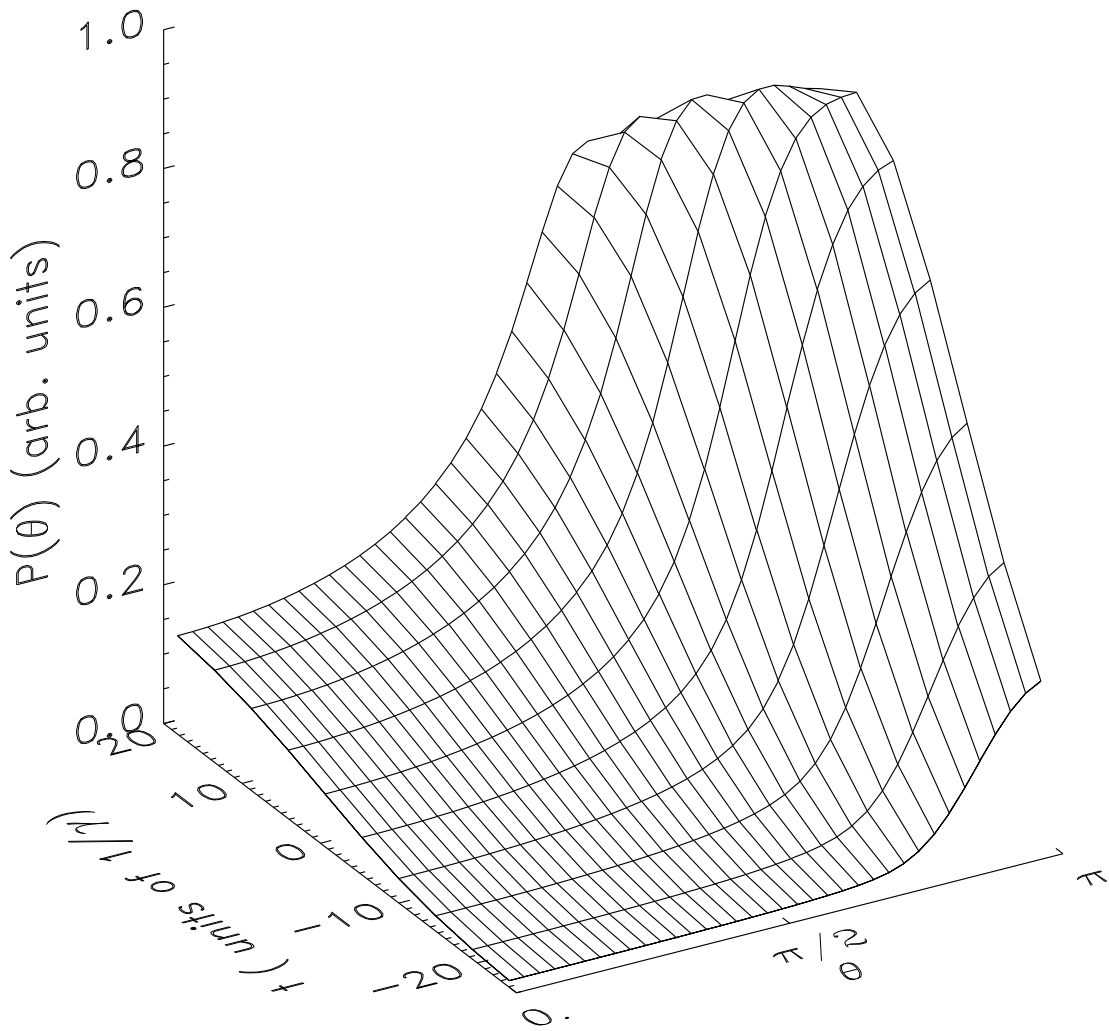